\documentclass[superscriptaddress,aps,preprint]{revtex4}
\usepackage[english]{babel}
\usepackage[latin1]{inputenc}
\usepackage{graphicx,color}
\usepackage{latexsym}
\usepackage{amsmath}
\usepackage{amsfonts}
\usepackage{amssymb}
\usepackage[latin1]{inputenc}

\def\bb{\bibitem}

\def\bb{\bibitem}

\newcommand{\be}{\begin{equation}}
\newcommand{\ee}{\end{equation}}
\newcommand{\bea}{\begin{eqnarray}}
\newcommand{\eea}{\end{eqnarray}}

\newcommand{\I}{\mathrm{i}}
\newcommand{\E}{\mathrm{e}}

\begin{document}

\title{Lorentz symmetry violation and an analog of Landau levels}

\author{E. Passos}
\email{passos,lrr,furtado,jroberto@fisica.ufpb.br}
\affiliation{{ Departamento de
F\'{\i}sica, Universidade Federal da Para\'\i ba, Caixa Postal 5008, 58051-970,
Jo\~ao Pessoa, PB, Brazil}}

\author{L. R. Ribeiro}
\email{passos,lrr,furtado,jroberto@fisica.ufpb.br}
\affiliation{{ Departamento de
F\'{\i}sica, Universidade Federal da Para\'\i ba, Caixa Postal 5008, 58051-970,
Jo\~ao Pessoa, PB, Brazil}}

\author{C. Furtado}
\email{passos,lrr,furtado,jroberto@fisica.ufpb.br}
\affiliation{{ Departamento de
F\'{\i}sica, Universidade Federal da Para\'\i ba, Caixa Postal 5008, 58051-970,
Jo\~ao Pessoa, PB, Brazil}}

\author{J. R. Nascimento}
\email{passos,lrr,furtado,jroberto@fisica.ufpb.br}
\affiliation{{ Departamento de
F\'{\i}sica, Universidade Federal da Para\'\i ba, Caixa Postal 5008, 58051-970,
Jo\~ao Pessoa, PB, Brazil}}
\affiliation{Instituto de F\'\i sica, Universidade de S\~ao Paulo\\
Caixa Postal 66318, 05315-970, S\~ao Paulo, SP, Brazil}

\begin{abstract}
Within the context of Lorentz violating extended electrodynamics,
we study an analog of Landau quantization for a system where a neutral particle moves in the presence of an electromagnetic field and a constant four-vector that breaks Lorentz symmetry. The nonrelativistic Hamiltonian associated to this system is obtained using 
the Foldy-Wouthuysen transformation for a Dirac spinor. The degenerated energy spectrum is obtained for a time-like and a space-like parameter Lorentz-breaking vector. The energy dependence of the cyclotron rotation direction in terms of supersymmetric quantum mechanics is analyzed.  
\end{abstract}

\maketitle
\section{Introduction}    
Lorentz and CPT symmetries are fundamental ingredients in the standard model of particle physics. A considerable number of recent studies investigating the possibility of violation of the Lorentz symmetry have been realized. The physical results found within these studies are the discovery of the birefringence of the light in the vacuum and the rotation of the plane of polarization of light \cite{jk}. The original motivation to study Lorentz symmetry
breaking emerged in the context of string theory \cite{kost}. A standard model extension including all possible terms which may violate Lorentz and/or CPT invariance, was presented in Ref. \cite{LV}. Noncommutative field theories offer another field-theoretic context for Lorentz violation within which one realistic models form a subclass of the standard model extending models \cite{NC}.

Recent studies addressed the analysis of the dynamics of a charged fermionic particle in an external electromagnetic field within the framework of the extended electrodynamics which includes the Lorentz-breaking terms, have been done by several authors \cite{LVQM1,LECJ,LVQM2,LVQM3}. As a further development of these studies, we intend to investigate the dynamics of a neutral nonpolarized spin-$\frac12$ particle in the presence of an electromagnetic field, taking into account the influence of a constant parameter that controls the Lorentz-symmetry violation. To be more precise, we are planning to obtain the Landau levels for such a particle. Basing on the idea of the standard model extension, we start with a model in which a Chern-Simons-like term \cite{jk} is added to the Dirac equation, and such term involves a background that characterizes the preferred direction in the space-time. In the sequel, we consider the nonrelativistic Hamiltonian associated to this system \cite{LECJ}, which was obtained using the Foldy-Wouthuysen (FW) transformation \cite{fw} for the Dirac spinors. 

The central aim of this paper is to obtain a quantization for a neutral particle energy levels in the system described above, which are treated as analogs of the standard Landau levels \cite{landau}, where a charged particle moves in the presence of a homogeneous magnetic field. We adapt the idea formulated in \cite{ericsson} to our system and choose a field-particle configuration where the analogs of Landau levels take place. The interest in the study of Landau quantization comes from the possibility to describe many problems such as the quantum Hall effect \cite{prange}, different two-dimensional surfaces \cite{comtet,dunne}, anyons excitations in rotating Bose--Einstein condensates \cite{paredes1,paredes2}, analog of Landau levels for polarized neutral particles \cite{ericsson,pla1,pla2} and other systems.

This paper is organized as follows: In the section II, we present the model and the nonrelativistic associated Hamiltonian, obtained via Foldy-Wouthuysen (FW) transformation. In the section III, we obtain the energy and angular momentum eigenvalues, as well as the associated eigenfunctions, for the space-like and time-like components of the Lorentz-symmetry violation parameter. In the section IV, we explain the energy dependence in the cyclotron rotation direction in terms of supersymmetric quantum mechanics \cite{ericsson,roy}. Finally, in the section V, we present our conclusions and remarks.


\section{Quantum Dynamics and Lorentz Symmetry Breaking}
The dynamics of a single neutral spin-$\frac12$ particle moving in the presence of the electromagnetic field which involves a Lorentz-breaking parameter is described by Dirac equation (we use the natural units system $\hbar=c=1$),
\begin{equation}
\Bigl(\I\gamma_\mu\partial^\mu+g\upsilon_{\alpha} {^*}F^{\alpha\beta}\gamma_{\beta}-m\Bigl)\psi=0\;,\label{01}
\end{equation}
where $g$ is a coupling constant, $^*F^{\mu\nu}=\frac12\epsilon^{\mu\nu\rho\sigma}F_{\rho\sigma}$ is the usual dual electromagnetic stress tensor 
where $\upsilon^{\mu}$ is a constant parameter implementing Lorentz symmetry violation, with $\upsilon_{0}$ and $\vec{\upsilon}$ being the time-like and space-like components of the coefficient $\upsilon_{\mu}$ respectively, and the Dirac matrices are given by:
\bea
\hat{\beta}=\gamma^{0}=\left(\begin{array}{cc}
1 & 0 \\
0 & -1\\ \end{array}\right)\;,\quad\gamma^{j}=\left(\begin{array}{cc}
0 & \sigma^{j} \\
-\sigma^{j} & 0\\ \end{array}\right)\;\;,\;\;\, \alpha^{i}=\hat{\beta}\gamma^{i}
\eea
where $\sigma_{i}$ are the Pauli matrices satisfying the  following algebra:
\bea
&&[\gamma^{0}\gamma^{i}-\gamma^{i}\gamma^{0}]=2\I\sigma^{0i},\nonumber\\&&[\gamma^{i}\gamma^{j}-\gamma^{j}\gamma^{i}]=2\I\sigma^{ij}=\I\epsilon^{ijk}\Sigma_{k},\nonumber\\&&\{\gamma^{i}\gamma^{j}+\gamma^{j}\gamma^{i}\}=2\delta^{ij}.
\eea 
Now, we can write the Eq. (\ref{01}) in the form
\be\label{2}
\Bigl[\I\gamma^\mu\partial_\mu - g\upsilon_{0}(\vec{\gamma}\cdot\vec{B})+g\vec{\gamma}\cdot(\vec{\upsilon}\times\vec{E})+(\vec{\upsilon}\cdot\vec{B})\gamma^{0}-m\Bigl]\psi=0\;
\ee
or, as is the same,
\bea\label{3}
H\psi=\bigl[\vec{\alpha}\cdot\vec{\pi}
-g\,(\vec{\upsilon}\cdot\vec{B})+\hat{\beta}m\bigl]\psi = \I\frac{\partial \psi}{\partial t},
\eea
where $\vec{\pi}=-\I\big[\vec{\nabla}+ g\bigl((\vec{\upsilon}\times\vec{E})-\,\upsilon_{0}\vec{B}\bigl)\big]$. The nonrelativistic Hamiltonian associated to Eq.(\ref{3}) is obtained by using the Foldy-Wouthuysen approximation. Following Ref. \cite{LECJ}, one can write down the associated nonrelativistic Hamiltonian in the form
\be\label{ham}
H= -\frac{1}{2m}\Bigl(\vec{\nabla}-\I\vec{a}\Bigl)^{2}+a_{0}\;.
\ee
 The expression (\ref{ham}) is similar to the Hamiltonian describing interaction of a particle minimally coupled to a constant Abelian gauge field with potential $a_{\mu}$, with  electric and magnetic fields. The vector potential $\vec a$ is written as 
\be
\vec{a}=g\vec{\upsilon}\times\vec{E}-g\upsilon_{0}\vec{B}\;, 
\ee
and the scalar potential looks like
\be a_{0}=\frac{g}{2m}\vec{\sigma}\cdot\big[\vec{\nabla}\times(\vec{\upsilon}\times\vec{E})\big]-\frac{g\upsilon_{0}}{2m}\vec{\sigma}\cdot\big[\vec{\nabla}\times\vec{B}\big]
+g\vec{\upsilon}\cdot\vec{B}\;.
\ee

\section{Landau analog levels}

At this point, we want to use the nonrelativistic Hamiltonian given by Eq. (\ref{ham}) to the study of analogs of Landau levels for neutral particles in the presence of electromagnetic field and a constant four-vector that breaks the Lorentz-symmetry. Note that the Eq. (\ref{ham}) is similar to the Hamiltonian of a charged particle in the presence of a external magnetic field. In this way, we can calculate the quantized energy levels in a analog way to Landau quantization. Thus, we can determine some aspects of the nature of the Lorentz-violating background by investigating the coupling constants properties. Neglecting the second order terms, we can write (\ref{ham}) in the form
\begin{eqnarray}
	H&=&\frac{1}{2m}\left[\vec p-g\vec\upsilon\times\vec E+ g\upsilon_0\vec{B}\right]^2
	+\frac{g}{2m}\vec{\sigma}\cdot\big[\vec{\nabla}\times(\vec{\upsilon}\times\vec{E})\big] \nonumber\\
	&&-\frac{g\upsilon_{0}}{2m}\vec{\sigma}\cdot\big[\vec{\nabla}\times\vec{B}\big]\;,
	\label{eq:hamold}
\end{eqnarray}
where $\vec p=-\I\vec{\nabla}$. We may assume that the fields are defined in the $x$-$y$ plane, hence we have only the $\sigma_3$ component of the spin vector in the Eq. (\ref{eq:hamold}). However, since our interest consists in the study of some aspects of nonrelativistic quantum mechanics, we restrict ourselves to the spin-up component only. Then, we rewrite the Hamiltonian (\ref{eq:hamold}) as 
\begin{eqnarray}
	H&=&\frac{1}{2m}\left[\vec p-g\vec\upsilon\times\vec E+ g\upsilon_0\vec{B}\right]^2
	+\frac{g}{2m}\vec{k}\cdot\big[\vec{\nabla}\times(\vec{\upsilon}\times\vec{E})\big] \nonumber\\
	&&-\frac{g\upsilon_{0}}{2m}\vec{k}\cdot\big[\vec{\nabla}\times\vec{B}\big]\;,
	\label{eq:ham2}
\end{eqnarray}
where $\vec{k}=(0,0,1)$ is a unitary vector directed along the $z$-axis.
 
From now on, we consider the constant 4-vector of the form $\upsilon^\mu=(\upsilon_0,\upsilon^i)$ where $\upsilon_0$ is the time-like component and $\upsilon^i$ are the space-like components.  In the next subsections  we study the quantum dynamics for both  time-like and space-like cases.

\subsection{Time-like Case}
If we consider the time-like terms only, we write the Hamiltonian as
\begin{equation}
	H=\frac{1}{2m}\left[\vec p+ g\upsilon_0\vec{B}\right]^2
	-\frac{g\upsilon_{0}}{2m}\vec{k}\cdot\big[\vec{\nabla}\times\vec{B}\big]\;.
	\label{eq:hamtl}
\end{equation}

Under a certain field configuration, neutral particles can display quantized energy levels in a way similar to the standard Landau levels \cite{landau} for charged particles. The precise system configuration conditions, under which an analog of Landau quantization occurs for neutral particles, are established in \cite{ericsson}. Here, we adapt these conditions to a system described by Eq. (\ref{eq:ham2}), where a neutral particle moves in the presence of a background that violates the Lorentz-symmetry. Hence, the torque over the particle must be zero, the fields must be static and the ``effective magnetic field'' $\vec{B}_\mathrm{eff}^{(\mathrm{t})}=\vec{\nabla}\times\vec{A}^{(\mathrm{t})}=\vec{\nabla}\times\vec B$ must be uniform. In the symmetric gauge, these conditions are satisfied, if the particle moves in the $x$-$y$ plane. So, let us choose the configuration
\begin{eqnarray}
	&\vec{k}=(0,0,1)\;,\nonumber\\[-3mm]\label{eq:eq09}\\[-3mm]
	&\vec A^{(\mathrm{t})}=\vec B=\dfrac{j}{2}(-y,x,0)\;,\nonumber	
\end{eqnarray}
where $j$ in a density of electric current. Thus, we may write the Hamiltonian (\ref{eq:hamtl}) as
\begin{eqnarray}
\hat{H}^{(\mathrm{t})}&=&\dfrac{1}{2m}\left[\hat{p}_x^2+\hat{p}_y^2+\left(\dfrac{m|\omega_\mathrm{t}|}{2}\right)^2(\hat{x}^2+\hat{y}^2)\right]\nonumber\\&&+\dfrac{\varsigma|\omega_\mathrm{t}|}{2}(\hat{L}_z^{(\mathrm{t})}-1)\;,
	\label{eq:10}
\end{eqnarray}
where $\varsigma|\omega_\mathrm{t}|=\frac{\varsigma|g\upsilon_0j|}{m}$ is the cyclotron frequency, $\varsigma=\pm$ labels the cyclotron rotation direction, and $\hat{L}_z^{(\mathrm{t})}=\hat{x}\hat{p_y}-\hat{y}\hat{p}_x$ is the angular momentum operator. In this case, the natural unit length is $\ell_\mathrm{t}=|g\upsilon_0j|^{-1/2}$. Since the coordinates and momenta obey the commutation relations
\begin{equation}
	[\hat{x},\hat{p}_x]=\I\;,\quad[\hat{y},\hat{p}_y]=\I\;,\quad[\hat{x},\hat{p}_y]=[\hat{y},\hat{p}_x]=0\;,
	\label{eq:11}
\end{equation} 
we define the ladder operators
\begin{eqnarray}
	\hat{a}&=&\dfrac{1}{\sqrt{2}}\left[\dfrac{\sqrt{m|\omega_\mathrm{t}|}}{2}(-\varsigma\hat x+\I\hat y)+\dfrac{(-\I\varsigma\hat{p}_x-\hat{p}_y)}{\sqrt{m|\omega_\mathrm{t}|}}\right]\;,\nonumber\\[-2.5mm]
	\label{eq:12}\\[-2.5mm]
	\hat{a}^\dag&=&\dfrac{1}{\sqrt{2}}\left[\dfrac{\sqrt{m|\omega_\mathrm{t}|}}{2}(-\varsigma\hat x-\I\hat y)+\dfrac{(\I\varsigma\hat{p}_x-\hat{p}_y)}{\sqrt{m|\omega_\mathrm{t}|}}\right]\;,\nonumber
\end{eqnarray}
which satisfy the relation $[\hat{a},\hat{a}^\dag]=1$. Thus, we can write the Hamiltonian (\ref{eq:10}) in the compact form
\begin{equation}
	\hat{H}^{(\mathrm{t})}=|\omega_\mathrm{t}|\left[\hat{a}^\dag a+\frac{1}{2}(1-\varsigma)\right]\;.
	\label{eq:13}
\end{equation}
from which we obtain the analog of Landau energy spectrum for the time-like case
\begin{equation}
	E_N^{(\mathrm{t})}=\left[N+\frac{1}{2}(1-\varsigma)\right]|\omega_\mathrm{t}|\;,\qquad N=0,1,2,\dots
	\label{eq:14}
\end{equation}
To obtain the degenerated spectrum, we introduce the associated ladder operators
\begin{eqnarray}
	\hat{b}&=&\dfrac{1}{\sqrt{2}}\left[\dfrac{\sqrt{m|\omega_\mathrm{t}|}}{2}(-\varsigma\hat x-\I\hat y)+\dfrac{(-\I\varsigma\hat{p}_x+\hat{p}_y)}{\sqrt{m|\omega_\mathrm{t}|}}\right]\;,\nonumber\\[-2.5mm]
	\label{eq:15}\\[-2.5mm]
	\hat{b}^\dag&=&\dfrac{1}{\sqrt{2}}\left[\dfrac{\sqrt{m|\omega_\mathrm{t}|}}{2}(-\varsigma\hat x+\I\hat y)+\dfrac{(\I\varsigma\hat{p}_x+\hat{p}_y)}{\sqrt{m|\omega_\mathrm{t}|}}\right]\;,\nonumber
\end{eqnarray}
with $[\hat{b},\hat{b}^\dag]=1$ and
\begin{equation}
	\hat{L}_z^{(\mathrm{t})}=\hat{a}^\dag a-b^\dag b\;,
	\label{eq:16}
\end{equation}
and $m_\ell=(N-n)$ is the eigenvalue of $\hat{L}_z^{(\mathrm{t})}$, where $n$ is the eigenvalue of $b^\dag b$. In this way, we write the analog of degenerated Landau energy levels
\begin{equation}
	E_{n,m_\ell}^{(\mathrm{t})}=\left[n+m_\ell+\frac{1}{2}(1-\varsigma)\right]|\omega_\mathrm{t}|\;,
	\label{eq:17}
\end{equation}
where $m_\ell$ explicits the degenerescence of the states.

The eigenfunctions for the ground state in the Cartesian representation, $\Psi_{0,0}^{(\mathrm{t})}(x,y)=\langle x,y|0,0\rangle$, can be obtained using the annihilation operators properties
\begin{eqnarray}
	\langle x,y|\hat{a}|0,0\rangle&=&0\;,\nonumber\\[-2.5mm]\label{eq:17.1}\\[-2.5mm]
	\langle x,y|\hat{b}|0,0\rangle&=&0\;.\nonumber
\end{eqnarray}
Solving the differential equations system (\ref{eq:17.1}), we obtain the normalized ground state eigenfunction
\begin{equation}
	\Psi_{0,0}^{(\mathrm{t})}(x,y)=\frac{1}{\sqrt{2\pi}\ell_\mathrm{t}}\E^{-\frac{x^2+y^2}{4\ell_\mathrm{t}^2}}\;.
	\label{eq:17.2}
\end{equation}
The high excited states may be calculated with applying the creation operators, 
\begin{eqnarray}
|N,0\rangle&=&(N!)^{-1/2}(\hat{a}^\dag)^N|0,0\rangle\;,\nonumber\\[-3mm]\label{eq:17.21}\\[-3mm]
|N,n\rangle&=&(n!)^{-1/2}(\hat{b}^\dag)^n|N,0\rangle\;,\nonumber 
\end{eqnarray}
since $\Psi_{N,n}^{(\mathrm{t})}(x,y)=\langle x,y|N,n\rangle$, we write
\begin{equation}
	\Psi_{n,m_\ell}^{(\mathrm{t})}(x,y)=\frac{\langle x,y|(\hat{b}^\dag)^n(\hat{a}^\dag)^{n+m_\ell}|0,0\rangle}{\sqrt{(n+m_\ell)!n!}}\;,
	\label{eq:17.3}
\end{equation}
that is a set of eigenfunctions associated with the energy eigenvalue $E_N^{(\mathrm{t})}=[N+\frac12(1-\varsigma)]|\omega_\mathrm{t}|$.

\subsection{Space-like Case}
Now, considering the space-like terms only in the Hamiltonian (\ref{eq:ham2}) and choosing a preferential direction $\vec\upsilon=\upsilon_1\vec{n}$, where $\vec{n}$ is a unitary vector, we have
\begin{equation}
	H=\frac{1}{2m}\left[\vec p-g\upsilon_1\vec{n}\times\vec E\right]^2
	+\frac{g}{2m}\vec{k}\cdot\big[\vec{\nabla}\times(\vec{\upsilon}\times\vec{E})\big]\;.
	\label{eq:hamsl}
\end{equation}
In the same way as above, we also adapt the conditions for the existence of Landau analog quantization in the space-like case. The conditions for zero torque, static fields and uniformity of effective magnetic field $\vec{B}_\mathrm{eff}^{(\mathrm{s})}=\vec{\nabla}\times\vec{A}^{(\mathrm{s})}=\vec{\nabla}\times\vec{n}\times\vec{E}$ are satisfied if we choose the symmetric gauge configuration
\begin{eqnarray}
	&\vec{k}=(0,0,1)\;,\quad\nonumber\\[-3mm]\label{eq:eq18}\\[-3mm]
	&\vec{A}^{(\mathrm{s})}=\vec{n}\times\vec{E}=\dfrac{\rho_\mathrm{e}}{2}(-y,x,0)\;,\nonumber	
\end{eqnarray}
where $\rho_\mathrm{e}$ is the density of electric charges, and the particle moves in the $x$-$y$ plane. Thus, the Hamiltonian (\ref{eq:hamsl}) can be written in the form
\begin{eqnarray}	H^{(\mathrm{s})}&=&\dfrac{1}{2m}\left[\hat{p}_x^2+\hat{p}_y^2+\left(\dfrac{m|\omega_\mathrm{s}|}{2}\right)^2(\hat{x}^2+\hat{y}^2)\right]\nonumber\\&&-\dfrac{\varsigma|\omega_{\mathrm{s}}|}{2}(\hat{L}_z^{(\mathrm{s})}-1)\;,
	\label{eq:18}
\end{eqnarray}
where $\varsigma|\omega_\mathrm{s}|=\frac{\varsigma|g\upsilon_1\rho_\mathrm{e}|}{m}$ is the cyclotron frequency, and $\hat{L}_z^{(\mathrm{s})}=\hat{x}\hat{p_y}-\hat{y}\hat{p}_x$ is the angular momentum operator. Here, the natural unit length is $\ell_\mathrm{s}=|g\upsilon_1\rho_\mathrm{e}|^{-1/2}$.The coordinates and momenta obey the commutation relation (\ref{eq:11}), and we may define the ladder operators and associated ladder operators
\begin{eqnarray}
	\hat{a}&=&\dfrac{1}{\sqrt{2}}\left[\dfrac{\sqrt{m|\omega_\mathrm{s}|}}{2}(\varsigma\hat x+\I\hat y)+\dfrac{(\I\varsigma\hat{p}_x-\hat{p}_y)}{\sqrt{m|\omega_\mathrm{s}|}}\right]\;,\nonumber\\
	\hat{a}^\dag&=&\dfrac{1}{\sqrt{2}}\left[\dfrac{\sqrt{m|\omega_\mathrm{s}|}}{2}(\varsigma\hat x-\I\hat y)+\dfrac{(-\I\varsigma\hat{p}_x-\hat{p}_y)}{\sqrt{m|\omega_\mathrm{s}|}}\right]\;,\nonumber\\[-2.5mm]
	\label{eq:19}\\[-2.5mm]
	\hat{b}&=&\dfrac{1}{\sqrt{2}}\left[\dfrac{\sqrt{m|\omega_\mathrm{s}|}}{2}(\varsigma\hat x-\I\hat y)+\dfrac{(\I\varsigma\hat{p}_x+\hat{p}_y)}{\sqrt{m|\omega_\mathrm{s}|}}\right]\;,\nonumber\\
	\hat{b}^\dag&=&\dfrac{1}{\sqrt{2}}\left[\dfrac{\sqrt{m|\omega_\mathrm{s}|}}{2}(\varsigma\hat x+\I\hat y)+\dfrac{(-\I\varsigma\hat{p}_x+\hat{p}_y)}{\sqrt{m|\omega_\mathrm{s}|}}\right]\;,\nonumber
\end{eqnarray}
which obey the commutation relations $[\hat{a},\hat{a}^\dag]=1$, $[\hat{b},\hat{b}^\dag]=1$. Thus we write the Hamiltonian and angular momentum operators in the compact form
\begin{eqnarray}
	\hat{H}^{(\mathrm{s})}&=&|\omega_\mathrm{s}|\left[\hat{a}^\dag\hat{a}+\frac12(1+\varsigma)\right]\;,
	\label{eq:20}\\
	\hat{L}_z^{(\mathrm{s})}&=&\hat{b}^\dag\hat{b}-\hat{a}^\dag\hat{a}\;.\label{eq:21}
\end{eqnarray}
Now, we obtain the energy and angular momentum spectrum
\begin{eqnarray}
	E_N^{(\mathrm{s})}&=&\left[N+\frac12(1+\varsigma)\right]|\omega_\mathrm{s}|\;,
	\label{eq:22}\\
	m_\ell&=&n-N\label{eq:23}\;.
\end{eqnarray}
Hence, since $N=n-m_\ell$, we write down the explicit degenerecences of energy levels,
\begin{equation}
	E_{n,m_\ell}^{(\mathrm{s})}=\left[n-m_\ell+\frac12(1+\varsigma)\right]|\omega_\mathrm{s}|\;.
	\label{eq:24}
\end{equation}
The ground state eigenfunction can be calculated, in the same way as in time-like case, using the properties of annihilation operators (\ref{eq:17.1}) and solving a differential equations system. Thus, we write the ground state normalized eigenfunction for the space-like case as
\begin{equation}
	\Psi_{0,0}^{(\mathrm{s})}(x,y)=\frac{1}{\sqrt{2\pi}\ell_\mathrm{s}}\E^{-\frac{x^2+y^2}{4\ell_\mathrm{s}^2}}\;,
	\label{eq:25}
\end{equation}
and using the creations properties we can obtain the high excited states
\begin{equation}
	\Psi_{n,m_\ell}^{(\mathrm{s})}(x,y)=\frac{\langle x,y|(\hat{b}^\dag)^n(\hat{a}^\dag)^{n+m_\ell}|0,0\rangle}{\sqrt{(n+m_\ell)!n!}}\;,
	\label{eq:26}
\end{equation}
which give a set of eigenstates associated to the energy level $E_N^{(\mathrm{s})}=[N+\frac12(1+\varsigma)]|\omega_\mathrm{s}|$.

We may treate the time-like coupling constant $g\upsilon_0$ as a kind of ``charge'' generated by the Lorentz-symmetry violation background. Thus identifying $q=g\upsilon_0$, where $q$ is the electric charge, we found that the behavior in this case is dual to the standard Landau levels \cite{landau} where a charged particle moves in the presence of a external homogeneous magnetic field, and presents quantized energy levels. In the same way, if we treate the space-like coupling constant as a magnetic polarization induced by the Lorentz-symmetry violation background, and identify $\mu=g\upsilon_1$, we have a system dual to the Landau--Aharonov--Casher levels \cite{ericsson} where a neutral magnetic polarized particle moves in the presence of a external electric field and presents a quantized energy spectrum.

\section{Supersymmetric Quantum Mechanics}

The dependence of the  energy levels on cyclotron rotation direction  may be understood in terms of supersymmetry \cite{ericsson,roy}. We suggest that the label $\varsigma=\pm$ is the eigenvalue of a operator $\hat{\tau}$ and introduce the supercharge
\begin{equation}
	\hat{Q}=\hat{a}\hat{f}^\dag\;,\quad\hat{Q}^\dag=\hat{f}\hat{a}^\dag\;,
	\label{eq:27}
\end{equation}
where $\hat{f}$ and $\hat{f}^\dag$ are the fermion annihilation and creation operators, and $\hat{a}$ and $\hat{a}^\dag$ are the boson annihilation and creation operators. Thus, $\hat{Q}$ creates a fermion and destroys a boson, and $\hat{Q}^\dag$ do the inverse operation. In the time-like case, the fermion ladder operators obey the relations $[\hat{f},\hat{f}^\dag]=\hat{\tau}$, $\{\hat{f},\hat{f}^\dag\}=1$, and $\hat{f}\hat{f}=\hat{f}^\dag\hat{f}^\dag=0$. So, we can write the Hamiltonian (\ref{eq:13}) in the form
\begin{equation}	\hat{H}^{(\mathrm{t})}=|\omega_\mathrm{t}|\left(\hat{Q}\hat{Q}^\dag+\hat{Q}^\dag\hat{Q}\right)=|\omega_\mathrm{t}|\left(\hat{a}^\dag\hat{a}+\hat{f}^\dag\hat{f}\right)\;.
	\label{eq:28}
\end{equation}
Therefore, we define the boson and fermion number operators as $\hat{a}^\dag\hat{a}$ and $\hat{f}^\dag\hat{f}=\frac12(1-\hat{\tau})$, and obtain the respective eigenvalues $N_\mathrm{B}=N$ and $N_\mathrm{F}=\frac12(1-\varsigma)$. In the space-like case, we define the fermions ladder operators satisfying the relations $[\hat{f},\hat{f}^\dag]=-\hat{\tau}$ and $\{\hat{f},\hat{f}^\dag\}=1$, where $\varsigma=\pm$ is the eigenvalue of the operator $\hat{\tau}$. In the same way, we can obtain the Hamiltonian for the space-like case as well as the boson number operator $\hat{a}^\dag\hat{a}$ and fermion number operator $\hat{f}^\dag\hat{f}=\frac12(1+\hat{\tau})$, and the respective eigenvalues $N_\mathrm{B}=N$ and $N_\mathrm{F}=\frac12(1+\varsigma)$.

\section{Concluding Remarks}
We investigated the possibility of Lorentz-symmetry violation in terms of a standard model extension. We carried out the study of a system where a single neutral spin-$\frac{1}{2}$ particle moves in the presence of the electromagnetic field and of the constant 4-vector that controls the Lorentz-symmetry violation. The Lorentz-symmetry breaking parameter can be understood as a background breaking the vacuum isotropy, in another words it may be viewed like a kind of ``aether''. In the nonrelativistic limit, we obtained a Hamiltonian similar in form to a system where a charged particle moves in the presence of a uniform magnetic field. Thus, imposing some field-particle configuration conditions, we obtained a quantized energy spectrum for the time-like and space-like cases, in the same way as within the the standard Landau quantization. We also showed that the dependence of energy levels  on the cyclotron rotation direction may be understood in terms of supersymmetric quantum mechanics. We proposed the interpretation of some aspects of the Lorentz-symmetry violation background supposing that it polarizes the particle, and the strength of the dipole may be measured by the coupling constants. Thus, in a hypothetical experiment, one may detect the Lorentz-symmetry violation parameter investigating changes in the particle polarization strength.

{\bf Acknowledgments.}We are indebted to Professor A. J. da Silva for the critical reading of the manuscript. This work was partially supported by the Funda\c c\~ao de Amparo \`a Pesquisa do Estado de S\~ao
Paulo (FAPESP), Con\-se\-lho Na\-cio\-nal de De\-sen\-vol\-vi\-men\-to Cient\'{\i}fico e Tecnol\'{o}gico (CNPq), CAPES/PROCAD  and Projeto Universal/CNPq.

\end{document}